\documentclass[bibyear]{aa}  
\usepackage{natbib}

\usepackage[colorlinks,breaklinks]{hyperref}

\newcommand{\sneae}{Type Ia supernovae}
\usepackage{graphicx}
\usepackage{txfonts}
\begin{document}

   \title{Peculiar velocity cosmology with type Ia supernovae\\[0.2em]
   \large CNRS/IN2P3 prospective white paper}
    
    \authorrunning{R. Graziani et al.}
    \titlerunning{IN2P3 prospective on peculiar velocity SN Cosmology}

   \author{R.~Graziani
          \inst{1},
          M.~Rigault
          \inst{1},
          N.~Regnault
          \inst{2},
          Ph.~Gris\inst{1},
          A.~Möller\inst{1},
          P.~Antilogus\inst{2},
        P.~Astier\inst{2},
        M. Betoule\inst{2},
        S. Bongard\inst{2},
        M.~Briday \inst{3},
        J. Cohen-Tanugi \inst{4},
          Y.~Copin\inst{3},
          H.~M.~Courtois\inst{3},
          D. Fouchez\inst{5},
          E.~Gangler\inst{1},
          D.~Guinet\inst{3},
          A. J. Hawken \inst{5},
          Y.-L.~Kim \inst{3},
          P.-F.~L\'eget \inst{2},
          J. Neveu \inst{6},
          P. Ntelis \inst{5},
          Ph. Rosnet \inst{1}
          E. Nuss \inst{4}
          }

   \institute{Universit\'e Clermont Auvergne, CNRS/IN2P3, Laboratoire de Physique
    de Clermont, F-63000 Clermont-Ferrand, France.
    \email{romain.graziani@clermont.in2p3.fr}
          \and 
          Laboratoire de Physique Nucl\'eaire et des Hautes Energies, Universit\'e Pierre et Marie Curie, Universit\'e Paris Diderot, CNRS/IN2P3, 4 place Jussieu, F-75005 Paris, France; Sorbonne Universités, UPMC Univ Paris 06, UMR 7585, LPNHE, F-75005, Paris, France.
          \and
          Universit\'e de Lyon, Univ. Claude Bernard Lyon 1, CNRS/IN2P3, IP2I Lyon, F-69622, Villeurbanne, France.
          \and
          Université de Montpellier, CNRS/IN2P3, Laboratoire Univers et Particules de Montpellier.
          \and
          Aix Marseille Univ, CNRS/IN2P3, CPPM, Marseille, France.
          \and 
          Université Paris-Saclay, CNRS/IN2P3, IJCLab, 91405 Orsay, France}
   \date{}

 \abstract{Type Ia Supernovae have yet again the opportunity to revolutionize the field of cosmology as the new generation of surveys are acquiring thousands of nearby SNeIa opening a new era in cosmology: the direct measurement of the growth of structure parametrized by $fD$. This method is based on the SNeIa peculiar velocities derived from the residual to the Hubble law as direct tracers of the full gravitational potential caused by large scale structure. With this technique, we could probe not only the properties of dark energy, but also the laws of gravity. In this paper we present the analytical framework and forecasts. We show that ZTF and LSST will be able to reach 5\% precision on $fD$ by 2027. Our analysis is not significantly sensitive to photo-typing, but known selection functions and spectroscopic redshifts are mandatory. We finally introduce an idea of a dedicated spectrograph that would get all the required information in addition to boost the efficiency to each SNeIa so that we could reach the 5\% precision within the first two years of LSST operation and the few percent level by the end of the survey.}

  \maketitle
%

\section{Introduction}
\label{sec:intro}
Luminosity distances derived from observations of Type Ia Supernovae (SNeIa) are a key tool to probe the expansion history of the Universe. They are particularly powerful in the late Universe ($z<0.3$) where dark energy is the driving component. Within the next 5 years, first the Zwicky Transient Facility (ZTF) and then the LSST will discover tens of thousands of nearby SNeIa over a large fraction of the visible sky, thanks to large field of view and fast cadences. 
This opens a new avenue for SN Cosmology: the direct measurement of the growth rate of structures in the recent Universe.

Residuals to the  Hubble law at low redshift are partially caused by the peculiar velocities of the SN host galaxies arising by their interaction with the underlying gravitational potential caused by large-scale structures in the universe. 
In practice, SN peculiar velocity leads to a difference between the observed redshift and the redshift due to the cosmological expansion of the universe. As SN brightness can be used to determine the latter, the observed redshift can effectively be used to constrain the line-of-sight peculiar velocity with respect to us. These measurements thus directly probe the total distribution of matter (including dark matter) unlike measurements of galaxy distribution only. The latter are sensitive to varying distributions for different matter types ; the so called "galaxy-bias".

The correlations of peculiar motions provide a direct measurement of the growth rate of structure, $fD$, which depends on both the Hubble expansion law and the nature of
gravity. The dependence upon the gravity model (general relativity or modified gravity) can be parametrized using the growth index $\gamma$, $fD=\Omega_m^\gamma$ \citep[e.g.,][]{linder2005}. Any deviation from $\gamma = 0.55$ (corresponding to a GR+$\Lambda$CDM model) would be an evidence for an extension to the standard model of cosmology. Interestingly, a first hint of physics beyond the standard model of cosmology has been seen when comparing the sole other directly measurable cosmological parameter: $H_0$~\citep{planck2018,riess2019}. Predictions from the $\Lambda{}$CDM model anchored on Planck (or WMAP) are indeed $\sim5\sigma$ lower than what is directly measured either by the direct distance ladder toward SNeIa~\citep{riess2019} or using time delays of strongly lensed quasar images~\citep{wong2019}. 

Growth rate measurements based on peculiar velocities are currently relying on galaxy scaling relations like the Tully-Fisher (TFR) and Fundamental Plane (FP) relations. These distance indicators lead to individual measurements 5 times less precise than SNeIa, but, until now, have been easier to gather in a large sample. The state-of-the-art 6dFGS~\citep{adams2017} uses FP data to measure $fD$ at the 15\% precision level at redshift $z=0$. Upcoming peculiar velocity surveys such as TAIPAN (\citealt{dacunha2017}, FP) and WALLABY (\citealt{johnston2008}, TFR) are expected to achieve a 7\% and 6.3\% measurement respectively, assuming no systematic error on the calibration of their distance indicators. The growth rate can also be measured using Redshift Space Distortions (e.g with DESI, 10\% measurement at $z\sim 0.3$) though in a redshift range where gravity models are hardly distinguishable. Moreover, this technique relies heavily on galaxy-bias models. We should also underline here that SN Ia is unique to probe the universe at low redshift (z<0.2) where dark energy is expected to dominate:  weak lensing and BAO, due to cosmic variance,  are “blind” in this late universe.

In this article we show that it is possible to achieve an unprecedentedly precise estimation of $fD$ free from galaxy-bias using SNeIa measurements.
We present the derivation of the growth rate of structure from SNeIa in section \ref{sec:fsigma8} and we describe the ZTF and LSST samples in section~\ref{sec:surveys}. We will base our forecast on a realistic scenario for both of these surveys and we will present our expectation in section~\ref{sec:forecasts}. More optimistic cases are also shown to illustrate the impact of additional facilities such as spectroscopic follow-up. 
Then in section~\ref{sec:discussion} we will discuss potential systematic effects in our analysis and the impact of missing spectroscopy (SNeIa typing and host redshift) and a way to solve these problems. We will
conclude in section~\ref{sec:conclusion}.


\section{Type Ia supernovae as a probe of growth of structure}
\label{sec:fsigma8}

On scales larger than individual galaxies, the peculiar motion $\mathbf{v}$ of a SN at a position $\mathbf{x}$ is driven by the large scale mass distribution surrounding the SN through the continuity equation:

\begin{equation}
    \nabla \cdot \mathbf{v}( \mathbf{x}, a ) = - a H(a) f(a) \delta (\mathbf{x},a), 
\end{equation}
where $a$ is the cosmological scale factor, $H$ the Hubble parameter, $\delta$ the density field and $f \equiv \frac{d\ln{D}}{d\ln{a}}$ is the linear growth rate. 

These velocity measurements thus directly sample the full gravitational density field induced by the combined effects of dark and baryonic matter, without any need for a galaxy-bias correction.

The peculiar velocities effectively induce correlated scatter along the redshift axis of the SN Hubble diagram. At low redshifts probed by ZTF and LSST, and assuming magnitude scatter (e.g.\ due to intrinsic magnitude dispersion) is small, this effect is pronounced. 

\subsection{Peculiar velocities as a probe of gravity}
The statistical distribution of peculiar velocities can be directly related to the natural laws that govern how primordial overdensities have evolved. 
The velocity-velocity power spectrum $P_{vv}$ is sensitive to the growth of structure as $P_{vv}\propto (fD\mu)^2$, where $D$ is  the spatially-independent ``growth factor'' in the linear evolution of density perturbations, $f \equiv \frac{d\ln{D}}{d\ln{a}}$ is the linear growth rate where $a$ is the scale factor   and $\mu\equiv \cos{(\hat{k} \cdot \hat{r})}$ where $\hat{r}$ is the direction of the line of sight and $\hat{k}$ is the direction of the wave-vector~\citep{hui2006,davis2011}.
As an example of how different theories of gravity affect the growth rate $f$; \citet{linder2005,linder2007} have found that General Relativity, $f(R)$, and DGP gravity follow the relation $f \approx \Omega_M^\gamma$ with the growth index $\gamma=0.55, 0.42, 0.68$ respectively \citep[see][for a review or these  models]{huterer2015}. Using this parameterization, peculiar velocity surveys probe gravity by modeling $fD=\Omega_M^{\gamma} \exp{\left(-\int_a^1 \Omega_M^{\gamma} d\ln{a} \right)}$, where $\Omega_M(a)$ also depends on the gravity model.
The  $\gamma$-dependence of $fD(z)$ is shown  in Figure~2 of  \citet{linder2013}. The parameter $\sigma_8$, the  standard deviation of overdensities in 8$h^{-1}$Mpc spheres, is commonly used in place of $D$ to normalize the overall amplitude of  overdensities, so the standard parameterization used by the community is $f\sigma_8$. However, the parametrization through $D(z)$ more precisely takes into account the redshift evolution of the correlations, and is more adapted for surveys exploring large redshift range like LSST. By definition $f\sigma_8\propto fD(z=0)$.

The same SNeIa used  to measure peculiar velocities can also serve as tracers of mass overdensities.  Overdensities and their motions are also connected by the continuity equation and the SN density-density power spectrum $P_{\delta \delta }$  depends on gravity  as $P_{\delta \delta }\propto (bD + fD\mu^2)^2$ where $b$ is the SN bias.  The bias is a ``nuisance'' parameter, not present in the velocity power spectrum, which must be marginalized out when inferring $fD$. Galaxy samples have traditionally been used to derive such Redshift Space Distortion (RSD) constraints on gravity. Constraints derived from SN~Ia samples will, however, be unique since the same field is responsible for both overdensity and velocity. When combined in a common analysis the sample variance limit is lowered.

Different approaches exist to model the velocity-velocity and velocity-density correlations (e.g.,~\citealt{lavaux2016,graziani2019,howlett2019}), and we adopt here a model-independant approach by focusing on Fisher-Matrix estimations. The precision in measuring $fD$ can be projected for different peculiar velocity surveys.
The primary  parameters that affect the precision are: solid angle $\Omega$, SN number density $n$, source intrinsic magnitude dispersion $\sigma_M$, and for a distance-limited survey the maximum distance $r_{\text{max}}$ (alternatively redshift $z_{\text{max}}$).
The dependence is most simply discerned 
in the  Fisher information matrix
\begin{align}
F_{ij} 
& = \frac{\Omega}{8\pi^2} \int_{r_{\rm min}}^{r_{\rm max}}  \int_{k_{\rm min}}^{k_{\rm max}}  \int_{-1}^{1} r^2 k^2 \text{Tr}\left[ C^{-1} \frac{\partial C}{\partial \lambda_i} C^{-1}
\frac{\partial C}{\partial \lambda_j} \right] d\mu\,dk\,dr
\label{eq:FM}
\end{align}
where
\begin{equation}
C(k,\mu,a)  =
  \begin{bmatrix}
   P_{\delta \delta}(k,\mu,a) + \frac{1}{n} &
   P_{v\delta}(k,\mu,a)  \\
   P_{v\delta}(k,\mu,a)  &
  P_{vv}(k,\mu,a) + \frac{\sigma^2}{n}
   \end{bmatrix},
\label{cov:eq}
\end{equation}
and the
peculiar-velocity  uncertainty  ($\sigma$) is related to magnitude uncertainty $\sigma_M$ and redshift uncertainty $\sigma_z$ via $\sigma^2 = (c\sigma_z)^2 + \left(\frac{\ln(10)}{5}\sigma_M H a \chi \right)^2$.
The dependence of $fD$ enters in the velocity-velocity correlation $P_{vv}\propto (fD\mu)^2$, the SN~Ia host-galaxy count overdensity power spectrum $P_{\delta \delta }\propto (bD )^2$, and the  galaxy-velocity cross-correlation $P_{v \delta} \propto  (bD)fD\mu$. The uncertainty in the growth rate is bounded by $\sigma_{fD}=\sqrt{F^{-1}_{fD,fD}}$.  Equation~\eqref{eq:FM} is used in section~\ref{sec:forecasts} to estimate forecasts for LSST and ZTF. We discuss in subsection~\ref{sec:vpeconlyornot} the interest of adding the density-velocity correlation into the growth rate estimation.

\section{Surveys}
\label{sec:surveys}
ZTF and the Wide-Fast-Deep survey of LSST are particularly suitable for peculiar velocity cosmology. They observe SNeIa on large scales, and with a known selection function. This latter point is critical in estimating velocity-density and density-density correlations. Their redshift ranges are within $z=0$ and $z=0.3$, where the cosmological information carried by peculiar velocities is maximum while individual observational errors remain relatively small. Furthermore, ZTF and LSST are complementary since their combined footprints cover the full sky. This section summarizes the main characteristics of the two surveys.

\subsection{ZTF Phase I and II}
\label{sec:ZTF}

The Zwicky Transient Facility \citep[ZTF,][]{bellm2019,graham2019} is an ongoing nearby universe survey observing the full northern sky every night in three bands with a $\sim20.5\,\mathrm{mag}$ (20) detection limit per exposure (i-band). See details in the aforementioned papers as well as in \cite{fremling2019}.

During its phase I (2018 Apr.~1 tp 2020 Dec.~31) ZTF will acquire about $\sim2,000$ spectroscopically typed SNeIa with $z<0.1$. Most of these transients ($\sim80$\%) are part of the completeness program that is classifying all extra-galactic transients reaching a brightness of 18.5~mag in any band \citep{fremling2019}. 
This trivial selection function means that the "completeness ZTF~Ia sample" is made of a random sampling of the vast majority of all SNeIa nature provides up to $z<0.08$ in the northern sky (18000 deg$^{2}$, Dec$>-30$, $|b|>5$).  By the end of phase~I, the "completeness ZTF Ia sample" should contain about 1600 SNeIa with typical 3-day cadence lightcurves in \textit{g-} and \textit{r-}band, half of which also having a typical 5 day cadence \textit{I}-band lightcurve. 

A 3 years extension of ZTF (named "Phase II") is currently been planned and might further focus on SNeIa cosmology, mainly for peculiar velocity analyses. Plans are made for ZTF Phase II to have a second dedicated spectroscopic follow up instrument installed on the 2.1m at Kitt Peak observatory. Together with the current SEDm \citep[60-inch at Palomar][]{rigault2019,blagorodnova2018}, ZTF will then be able to extend its completeness program down to 19~mag. The phase coverage in \textit{g-} and \textit{r-} should be reduced to a 2 day cadence and \textit{I-}band lightcurves would be acquired for every transient with a 4 day cadence. Altogether, this means that ZTF Phase II is expected to add about 7000 spectroscopically typed SNeIa with $z<0.12$, $\sim5000$ of which been part of the completeness sample now reaching $z<0.09$.

Phase I and Phase II will have high-quality spectroscopic redshifts either from public archives (a third) or from massive spectroscopic facilities such as DESI and WEAVE, with which ZTF is discussing agreements.

Throughout the rest of the paper, we will conservatively limit ourselves to the completeness ZTF~Ia sample with I-band ($\sim5500$ SNeIa), for which we already know how to extract the growth rate of structure signal with no further assumption. Shall ZTF be able to get $0.12\,\mathrm{mag}$ intrinsic scatter in 2 band only or perfectly know its selection function for higher redshift, we would be able to significantly increase the sample size and therefore the precision on deriving $fD$.

\subsection{LSST}
\label{sec:LSST}
The LSST will open up unprecedented new perspectives for supernovae and supernova cosmology through the observation of hundred of thousands of well-measured SNeIa. Only well-measured SNeIa will be used to constrain cosmological parameters. The size and quality of the sample will be governed by the observing strategy chosen by LSST. Recent studies \citep{lochner2018} have shown that it is possible to observe a complete sample of well-measured ($\sigma_\mu \leq 0.12$) SNeIa up to redshifts of 0.3-0.4 in the Wide-Fast-Deep survey. The SNeIa will be observed in four different bands $g$, $r$, $i$, $z$ plus some partial information in $u$ and $y$ bands. An accumulation of up to 300k \sneae~over $\sim$ 18,000$\deg^2$ can be obtained after ten years of operation by 
adjusting key observing strategy parameters such as the cadence, the depth, and the season length. About 10\% will benefit from a live spectrum ven though a dedicated spectroscopic facility can get one for all nearby $z<0.18$ SNeIa ; see Section~\ref{sec:addspectro}. From these 300k SNeIa, we assume here that only the ones at redshift $z<0.18$ will have a host galaxy redshift measurement (see subsection~\ref{sec:redshifts}) as the worst case scenario and the baseline for our predictions. This corresponds to $\sim8,000$ SNeIa per year. In section~\ref{sec:redshifts} we discuss a strategy to push up to $z<0.25$ by adding dedicated spectroscopic facilities (20,000~SNeIa per year). 

In contrast with ZTF, the fact that LSST SNeIa will be observed with at least four filters, including near infrared ones, could open the door for improved standardisation techniques, potentially reaching the $\sigma_\mu \sim 0.08$ \citep{fakhouri2015}. See the discussion Section~\ref{sec:addspectro} for the impact of such an improvement on the derivation of $fD$, see also fig.~\ref{fig:fD}.

\subsection{ZTF and LSST}
ZTF and LSST are located in each hemisphere and are therefore complementary for mapping the nearby Universe since they will,  together, probe structures over the full sky. The two surveys will overlap in time during the last year of ZTF Phase II providing LSST starts end 2022. During that time, they will share a declinaison band of $\sim20$\,deg which will enable us to accurately cross calibrated the instruments. Finally, while LSST goes deeper, it saturates at the brightest end of the Hubble Diagram (z<0.035) with the current observing strategy, ZTF will acquire those SNeIa and will anchor the very nearby flow of the Universe.

\section{Forecasts}
\label{sec:forecasts}
The expected relative precision on $fD$ for a survey can be estimated using the Fisher information Eq.~\eqref{eq:FM}. The Fisher information depends on the observed sky fraction $\Omega$, the redshift limit $z_{max}$, the linear smallest scale $k_{max}$, the uncertainty on the magnitude uncertainty $\sigma_M$ and the density of observations $n$. The main difference between LSST and ZTF is their redshift limits. The values used in this work are listed in table~\ref{tab:numbers}, we use $n=1.75 \times 10^{-5} \left(h/0.7\right)^3~\text{Mpc}^{-3}\text{yr}^{-1}$ and  $k_{max}=0.1\,h\text{Mpc}^{-1}$. 

Based on these simulations, we may develop forecasts on the fractional error on $fD$ for each survey using: (1) the full density-density + density-velocity + velocity-velocity correlations, and (2) peculiar velocities only. 
The latter contains less cosmological information but is less prone to systematics and selection effects issues.

\begin{table}[t]
    \caption{Parameters used for the Fisher-matrix forecasts showed in Figure~\ref{fig:fD}. We assume a constant volumetric density of SNeIa $n=1.75 \times 10^{-5} \left(h/0.7\right)^3~\text{Mpc}^{-3}\text{yr}^{-1}$ and we set the linear smallest scale $k_{max}=0.1\,h\text{Mpc}^{-1}$}
    \centering
    \begin{tabular}{r|ll}
        \hline \hline\\[-0.8em]
        Param & ZTF & LSST\\
         \hline\\[-0.8em]
        $z_{max}$ & 0.09 & 0.18 \\[0.10em]
        $\Omega$ & \multicolumn{2}{c}{$2.7\pi$} \\[0.10em]
        $\sigma_M$ & \multicolumn{2}{c}{$0.12\,\mathrm{mag}$}\\[0.10em]
        $\sigma_z$ & \multicolumn{2}{c}{$1.6 \times 10^{-4}$}\\[0.10em]
        \hline\\[-0.8em]
        N$_{\mathrm{SN}}$/yr & 1,000 & 8,000\\
        \hline \hline
        \end{tabular}
    \label{tab:numbers}
\end{table}

\begin{figure*}[h]
\centering
\includegraphics[width=1.\textwidth]{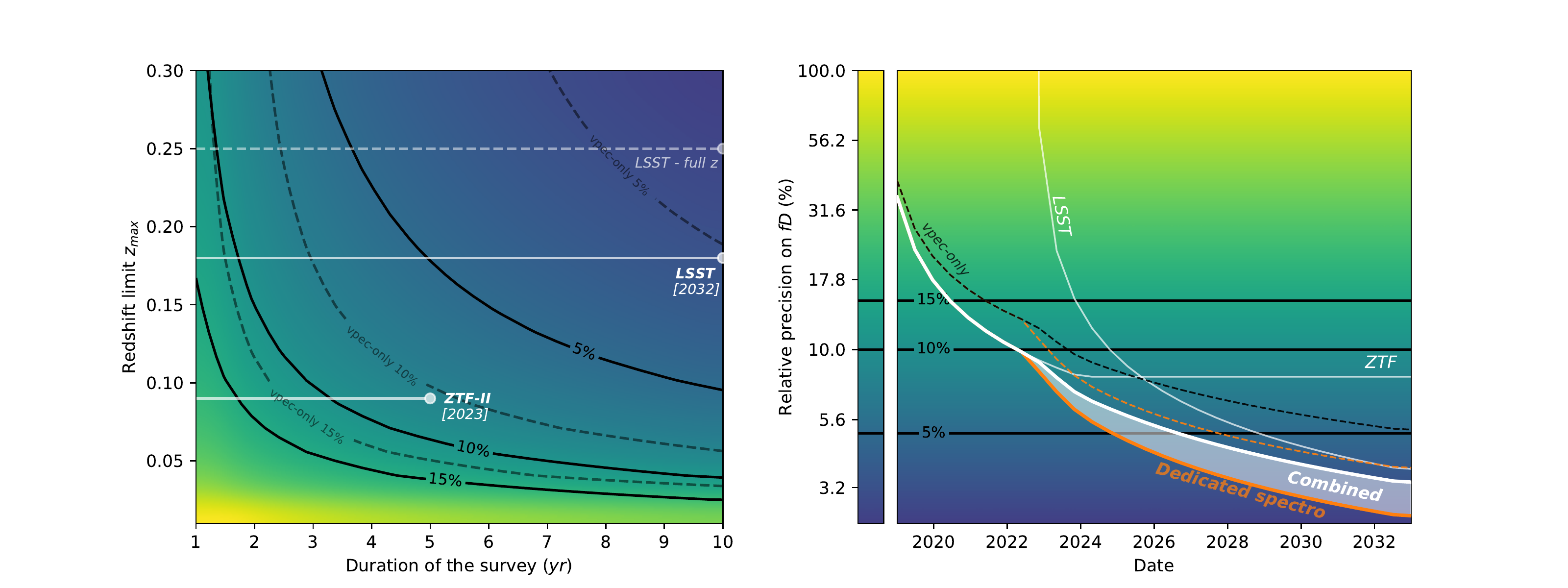}
\caption{(Left) Relative precision on the growth rate $fD$ for SNeIa surveys as a function of the redshift limit and the duration of the survey, assuming a constant density of object and a magnitude precision of 0.12. Black lines show the 15, 10 and 5\% limits on the precision for the full analysis (number counts and peculiar velocities). The corresponding dashed lines are for the peculiar velocity analysis only. Estimations for LSST and ZTF as a function of time are shown in white assuming redshift limits of 0.18 and 0.09 respectively. (Right) Relative precision on $fD$ for LSST, ZTF and the combination of the two, as a function of time. The thick white line shows the combination of LSST and ZTF assuming $\sigma_M=0.12\,\mathrm{mag}$. If LSST SNeIa were to get a better standardisation, the precision would be significantly improved, as shown by the white band. The orange line represents the $\sigma_M=0.08\,\mathrm{mag}$ that could be achieved with a dedicated spectrograph. The grey (resp. orange) dashed line represents the combined forecast when using the peculiar velocity only assuming  $\sigma_M=0.12\,\mathrm{mag}$ (resp. $\sigma_M=0.08\,\mathrm{mag}$ for  LSST SNeIa)}.\label{fig:fD}
\end{figure*}

The left panel of figure~\ref{fig:fD} shows the expected relative precision on $fD$ (color, in log scale) as a function of the redshift limit and the duration of the survey.
We present the 5, 10 and 15\% level contours for both the full densitiy-velocity and the velocity-only analyses and for both surveys.

Figure~\ref{fig:fD} illustrates the significant impact of the full analysis since LSST could reach 5\% precision on $fD$ in 5 years, whereas it would take 10 years for the velocity-only analysis.

Figure~\ref{fig:fD} also shows that adding density information has larger impact for higher redshifts. This is because the individual errors on peculiar velocities are small at low-z. The precision on $fD$ measurement is driven by velocity correlations. At higher redshift, however, the observational volume is large and the number of objects makes the velocity-density correlations play a major role. 
We discuss in section~\ref{sec:vpeconlyornot} that systematic effects due to the departure from linearity are to appear at the few percent precision.

The right panel of figure~\ref{fig:fD} shows the evolution of the $fD$ measurement as a function of time. As ZTF is already collecting data, the survey will lead the $fD$ cosmology up to the mid-2020, by then, LSST will take over. We highlight that the combination could lead to the first 5\% precision direct measurement of the growth of structure very weakly sensitive to the galaxy-bias in 2027. \textbf{This 5\% relative uncertainty on $fD$ is equivalent to an absolute error on $\gamma$ of 0.06, a precision at which DGP and $f(R)$ gravity models can be tested at the $2\sigma$ level}. For comparison, DESI is expected to measure $fD$ at the 10\% level.

\section{Discussion}
\label{sec:discussion}

The former forecast section only models statistical uncertainty on the measurement of the growth rate, assuming that: (i) SNeIa are perfectly typed, (ii) their host galaxy redshift are known and (iii) the selection function of the survey is perfectly known. In this section we discuss these three assumptions and their potential impact on our forecasts. We also discuss the possible systematics coming from the density-velocity cross-correlation model. 

\subsection{Caveats with density information}
\label{sec:vpeconlyornot}

The density field of matter drives both the location and the peculiar velocity of observed SNeIa. Accounting for the observed spatial distribution of SNe thus constrains the density field which, in turn, constrains the peculiar velocities. 
Mathematically, velocities are proportional to $fD$, hence, the velocity-velocity auto-correlation depends on $(fD)^2$. Velocity-density cross-correlation depends on $fD$ only. When all fitted together, the density-density term helps anchoring the velocity-density one at a cost of assuming the underlying connection between the SNeIa spatial distribution and the true matter distribution. 

In particular, the derivation of $fD$ assumes a linear evolution of the density field, valid for large scales only, while the spatial distribution of the SNeIa are following the local baryonic matter non-linear distribution. When adding the density information we therefore need to introduce two models that, if wrong, would bias our results: (1) how the spatial distribution of SNeIa probes the full matter distribution and (2) how the full matter distribution is connected to the large scale linear density distributions.

Analysis of redshift space distortions are based on galaxy spatial distribution only and are directly affected by the choice of these two models, so much so that constrains by BOSS~\citep{alam2017} on $fD$ include a $\sim 5\%$ systematic errors at redshift $z=0.38$ coming from the non-linear and bias models. In our case, this modelling issue is mitigated by the fact that peculiar velocities do not depend on the galaxy-bias and the velocity-velocity correlation directly constrain $fD$. Hence, while our objective is to reach a 5\% statistical level, the systematic uncertainties induced by the two aforementioned effects are significantly lowered when densities are cross-correlated with peculiar velocities and should not dominate our error budget. A detailed study of this effect will be presented in Graziani et al. in prep.

Forecasts using peculiar velocity only lead to conservative estimation of $fD$ but robust with respect to systematics. On the contrary, forecasts for the full analysis, \emph{i.e.} including velocity-density correlations, represent the best case scenarios for which controlling the systematic uncertainties  will require further investigations. Yet, Fig.~\ref{fig:fD} shows that the full analysis improves by 30\% the direct derivation of $fD$ in comparison to the peculiar velocity only analysis. When combined with ZTF, LSST SNeIa can reach the 5\% precision after 4 years of operation only (2026), while it would take 10 years if density is not included. In subsection~\ref{sec:addspectro}, we investigate a way of reaching this statistical precision with peculiar velocities only and to further improve the full analysis to reach the few percent precision of $fD$ within the LSST era.

\subsection{Typing}
\label{sec:typing}

Typing of supernovae is traditionally done using spectroscopy, which is rare, expensive and time critical since it has to be acquired within a time window of a couple of weeks around maximum of light. With the statistics of SNeIa discovered by the new generation of surveys, typing will not be possible for every SNe, and especially not for the most distant ones that are at the same time the most numerous and the faintest. Supernova classification based on photometric data only is a viable solution providing SN lightcurves are of sufficient quality for Ia photometric features to be identified by photo-typing algorithms. 
In the last few years, machine learning techniques have shown their strength in this field and could provide high-purity phototyping SNIa samples \citep{ishida2013,pasquet2019b,moller2020} especially when fed by scientifically motivated derived information \citep{boone2019}. It is therefore expected that one can obtain a photometrically typed SNeIa sample with a few percent non-Ia contamination during the LSST era. 
This might not be sufficient for classical Hubble diagram analyses given that the contaminating populations might evolve as a function of redshift in an uncontrolled way. For peculiar velocity studies that are made within a very limited redshift range, the first order impacts of non-Ia contamination simply are: (1) an increase of the sample magnitude scatter and (2) a larger fraction of peculiar velocity outliers. 

The first effect will be marginal given the expected few percents contamination (maybe increasing $\sigma_M$ to $0.13$\,mag). The second effect might further impact the analysis. 
Since a residual to the Hubble diagram is interpreted as a peculiar velocity, a non-Ia will be assigned a greatly exaggerated peculiar velocity in comparison to what it truly has. This could bias the measured velocity-velocity and velocity-density correlations and thereby the derived galaxy clustering. However, non-linear velocities have the same analytical signatures and most likely the fitter will consider this artificial high velocity as a sign of great non-linearity at the non-Ia location. It is to be expected that the non-linear modelling could be biased. This could in turn affects the linear part of the modelling from which the $fD$ measurements are derived. The effect of non-Ia contamination is therefore at most a second order effect, which should remain subdominant as long as the non-Ia contamination remains at the few percents level as one could expect. A  thorougher study of this systematic will be presented in Graziani et al. in prep.

We highlight here that ZTF will be fully spectroscopically typed (see section~\ref{sec:ZTF}) and that the sample this survey is gathering (full spectroscopic typing of all extra-galactic transients brighter than $18.5/19\,\mathrm{mag}$) will be true milestone for training future photo-typing techniques. Therefore, assuming percent contamination for the fraction of LSST transients non spectroscopically typed as we just did seems reasonable. We thus conclude that SN typing seems a marginal issue for our peculiar velocity analysis.


\subsection{Redshifts}
\label{sec:redshifts}

An error on the redshift measurement $\sigma_z$ of a host galaxy directly translates into an error on the observed peculiar velocity of $c \sigma_z$.
To illustrate the scale of the effect, an uncertainty of $\sigma_z = \mathcal{O}(10^{-3})$ would completely dominate the peculiar velocity signal which is expected to be around $300~\text{km.s}^{-1}$ for a $\Lambda$CDM model. Hence, to ensure an accurate measurement of $fD$, a redshift uncertainty $\mathcal{O}(10^{-4})$ is therefore needed.
Current state-of-the-art photo-redshift measurements optimized for nearby galaxies \citep{pasquet2019a} could reach a $\sigma_z\sim0.01$ plus a typical one percent rate of catastrophic event. It is therefore unlikely that the required level of precision for peculiar velocities could ever be achieve using photometric redshift estimations, and consequently peculiar velocity analysis need spectroscopic follow-up.

As discussed in section~\ref{sec:ZTF}, ZTF SNeIa will all have a spectroscopic redshift as it benefit from the SDSS' spectroscopic surveys, BOSS already observed approximately a third of the SN host and the rest will be acquired by DESI (most of the ZTF host are brighter than mag 20 and we have an agreement for the rest). 
LSST observing from the southern hemisphere, less spectroscopic facilities exist. However the 4MOST, in which IN2P3 is involved for this reason, has a dedicated program for acquiring SNeIa host redshift and, with other multi-fiber spectrograph such as TAIPAN, it is fair to expect that all nearby SNeIa ($z<0.18$) will have a spectroscopic redshift. 

\subsection{Additional spectroscopy}
\label{sec:addspectro}
To push further the analysis, we are considering building a dedicated facility half way between SNIFS and the SEDm in the south. This single object spectrograph should be a fully automatised facility installed in robotic 2m class telescope. This spectrograph should be a high efficiency IFU (slicer) such that (1) it is not sensitive to small pointing issue in order to maximize its efficiency as there is no need for pre-acquisition (2) it could do spectrophotometry and (3) it could acquire at the same time the host and the SN spectra.
The second point is of particular interest as recent SNIa analyses are pointing our that combining spectroscopic and photometric information could help significantly reduce the dispersion on the Hubble diagram. The twining technique is one example claiming to reach $\sigma_M = 0.07\,\mathrm{mag}$ \cite{fakhouri2015} but other standardisation methods such as SNEMO \citep{saunders2018} or SUGAR \citep{leget2019} are also claiming to significantly reduce the SN dispersion when spectroscopic information is added. It is therefore expected that combining spectral features with the (at least) 4 optical and near infrared bands of LSST, we could get at least $\sigma_M = 0.08\,\mathrm{mag}$.
The third point is also powerful for two reasons: first, the IFU strongly helps identifying the true host and therefore the good redshift in comparison to photometric analysis \citep{rigault2013,gupta2016}; second, getting the host spectra is a crucial element to account for astrophysical biases and thereby further reduce the SN dispersion \citep{sullivan2010,rigault2013,rigault2018}.

Hence, our proposed instrument could be able, on its own, to type and get a redshift of all $z<0.18$
SNeIa, especially when considering that  photo-typing can provide reasonable classification during the rising phase of the transient \citep{moller2020,muthukrishna2019}, which would limit non-Ia trigger. With such a facility in place, we would coordinate with 4MOST to avoid double triggering such that they could focus on larger redshift target with their 4m class telescope. 

This dedicated facility could be a game changer for peculiar velocities and we could reach the 5 \% precision on $fD$ by 2025 conservatively aiming at a $\sigma_M=008\,$mag dispersion (see fig.~\ref{fig:fD}, red line). Also, the velocity-only analysis, less prone to systematics and totally independent from the galaxy-bias, would reach the 5\% limit on 2028. It would in addition enable us to push the classical $0.12\,\mathrm{mag}$ dispersion study to $z<0.25$, which would by itself reduce the uncertainties on $fD$ by 15\%, i.e. 2 years of survey by 2030 ; see fig.~\ref{fig:fD}.


\subsection{Sample selection}
\label{sec:selection}

Because of two different effects, peculiar velocity analyses heavily rely on the knowledge and modelization of the survey sample selection. 

The first is the so-called Malmquist bias. It arises when a magnitude cut-off is applied to the data, discarding all the faintest objects. If not correctly modeled, it artifically shifts the average SNIa luminosity toward the bright end at a given distance, simply because the fainter ones are missing. In term of peculiar velocity analyses, this artificial negative average Hubble residual mimics the effect of a global inflow towards us, or, in another word, a spurious correlation on large scales. This falsifies the large scale structures of the nearby Universe and therefore the derivation of $fD$ ; as well as that of $H_0$. We have shown in~\citep{graziani2019} that this issue could be handle within our forward modeling framework, provided the selection function is known. 

The second effect is the accurate estimation of SNeIa density correlations. To compute these correlations, one typically needs to infer the underline density of object from the observed number counts, which always assumes that the probability distribution of observing an object at a specific point in space is perfectly known. Here again, it is mandatory to assume a known selection function.

These two effects explain why we are focusing on the "complete ZTF~Ia sample" and its trivial selection function, for this analysis. The $z<0.18$ LSST sample should have the same characteristics and we are actively working within the LSST DECS working groups to ensure so.

\section{Conclusion}
\label{sec:conclusion}

In this paper, we have presented that, thanks to the new generation of survey and the large statistics of SNeIa they are acquiring  at low redshift (z<0.3), a new probe is now accessible for Cosmology: the direct measurement of the growth of structure. This measurement can be used not only to test the properties of the energy contents of the Universe but also the theory of gravitation itself. 
By 2027, ZTF and LSST surveys will gather several thousands of SNeIa with $z<0.18$ which could enable us to derive $fD$ at the 5\% precision level. This would be sufficient to discriminate between General Relativity and alternative models of gravity and test if the direct measurement of the growth of structure is compatible with prediction based on Planck and weak lensing  in a similar fashion as how the direct measurement of $H_0$ could be compared with  $\Lambda{}\mathrm{CDM}$ predictions anchored by CMB data --~which happen to be in $5\sigma$ disagreement. We presented that for this analysis a good knowledge of the SNeIa selection function is necessary and while SN phototyping might be enough, having spectroscopic redshifts ($\sigma_z = \mathcal{O}(10^{-4}$)) is mandatory. In that context, we introduced the idea of developing a dedicated integral field spectrograph that could, simultaneously type, redshift and boost the precision of all the SNeIa LSST will discover up to $z<0.18$. With such facility, we could reach the 5\% precision in 3 years of LSST operation and we could aim at the few percent by the survey. Interestingly, the spectroscopic data could allow us to discard the SN location information in the derivation of $fD$ and still get a 5\% precision by 2028. Doing so will ensure that our analysis is free from all the major systematic effects such as the "baryon--dark matter" bias.

\begin{acknowledgements}
    This project is supported by the IN2P3 and has received funding from the European Research Council (ERC) under the European Union's Horizon 2020 research and innovation programme (grant agreement n 759194 - USNAC). Support has been provided by the Institut Universitaire de France, the CNES, and the region Auvergne-Rhone-Alpes.
    
    PN is funded by  'Centre National d'\'etude spatiale' (CNES), for the Euclid project.
\end{acknowledgements}

%

\begin{thebibliography}{} 
\bibitem[Adams \& Blake(2017)]{adams2017} Adams, C., \& Blake,
  C.\ 2017, \mnras, 471, 839

\bibitem[Alam et al.(2017)]{alam2017} Alam, S., Ata, M., Bailey, S., et al.\ 2017, \mnras, 470, 2617

\bibitem[Bellm et al.(2019)]{bellm2019} Bellm, E.~C., Kulkarni, S.~R., Graham, M.~J., et al.\ 2019, \pasp, 131, 018002

\bibitem[Blagorodnova et al.(2018)]{blagorodnova2018} Blagorodnova, N., Neill, J.~D., Walters, R., et al.\ 2018, \pasp, 130, 035003

\bibitem[Boone(2019)]{boone2019} Boone, K.\ 2019, \aj, 158,
  257

\bibitem[da Cunha et al.(2017)]{dacunha2017} da Cunha, E., Hopkins, A.~M., Colless, M., et al.\ 2017, \pasa, 34, e047

\bibitem[Davis et al.(2011)]{davis2011} Davis, T.~M., Hui, L., Frieman, J.~A., et al.\ 2011, \apj, 741, 67

\bibitem[Fakhouri et al.(2015)]{fakhouri2015} Fakhouri, H.~K., Boone, K., Aldering, G., et al.\ 2015, \apj, 815, 58

\bibitem[Fremling et al.(2019)]{fremling2019} Fremling, U.~C., Miller, A.~A., Sharma, Y., et al.\ 2019, arXiv e-prints, arXiv:1910.12973

\bibitem[Graham et al.(2019)]{graham2019} Graham, M.~J., Kulkarni, S.~R., Bellm, E.~C., et al.\ 2019, \pasp, 131, 078001

\bibitem[Graziani et al.(2019)]{graziani2019} Graziani, R., Courtois, H.~M., Lavaux, G., et al.\ 2019, \mnras, 488, 5438

\bibitem[Gupta et al.(2016)]{gupta2016} Gupta, R.~R., Kuhlmann, S., Kovacs, E., et al.\ 2016, \aj, 152, 154

\bibitem[Howlett(2019)]{howlett2019} Howlett, C.\ 2019, \mnras, 487, 5209

\bibitem[Hui \& Greene(2006)]{hui2006} Hui, L., \& Greene,
  P.~B.\ 2006, \prd, 73, 123526

\bibitem[Huterer et al.(2015)]{huterer2015} Huterer, D., Kirkby, D., Bean, R., et al.\ 2015, Astroparticle Physics, 63, 23

\bibitem[Ishida \& de Souza(2013)]{ishida2013} Ishida, E.~E.~O., \& de Souza, R.~S.\ 2013, \mnras, 430, 509

\bibitem[Johnston et al.(2008)]{johnston2008} Johnston, S., Taylor, R., Bailes, M., et al.\ 2008, Experimental Astronomy, 22, 151

\bibitem[Lavaux(2016)]{lavaux2016} Lavaux, G.\ 2016, \mnras, 457, 172

\bibitem[Linder(2005)]{linder2005} Linder, E.~V.\ 2005, \prd, 72, 043529

\bibitem[Linder(2013)]{linder2013} Linder, E.~V.\ 2013, \jcap, 2013, 031

\bibitem[Linder \& Cahn(2007)]{linder2007} Linder, E.~V., \& Cahn, R.~N.\ 2007, Astroparticle Physics, 28, 481

\bibitem[Lochner et al.(2018)]{lochner2018} Lochner, M., Scolnic, D.~M., Awan, H., et al.\ 2018, arXiv e-prints, arXiv:1812.00515

\bibitem[L{\'e}get et al.(2019)]{leget2019} L{\'e}get, P.-F., Gangler, E., Mondon, F., et al.\ 2019, arXiv e-prints, arXiv:1909.11239

\bibitem[M{\"o}ller \& de Boissi{\`e}re(2020)]{moller2020} M{\"o}ller, A., \& de Boissi{\`e}re, T.\ 2020, \mnras, 491, 4277

\bibitem[Muthukrishna et al.(2019)]{muthukrishna2019} Muthukrishna, D., Narayan, G., Mandel, K.~S., et al.\ 2019, \pasp, 131, 118002

\bibitem[Pasquet et al.(2019a)]{pasquet2019a} Pasquet, J., Bertin, E., Treyer, M., et al.\ 2019, \aap, 621, A26

\bibitem[Pasquet et al.(2019)]{pasquet2019b} Pasquet, J., Pasquet, J., Chaumont, M., et al.\ 2019, \aap, 627, A21

\bibitem[Planck Collaboration et al.(2018)]{planck2018} Planck Collaboration, Aghanim, N., Akrami, Y., et al.\ 2018, arXiv e-prints, arXiv:1807.06209

\bibitem[Riess et al.(2019)]{riess2019} Riess, A.~G.,
  Casertano, S., Yuan, W., et al.\ 2019, \apj, 876, 85

\bibitem[Rigault et al.(2013)]{rigault2013} Rigault, M., Copin, Y., Aldering, G., et al.\ 2013, \aap, 560, A66

\bibitem[Rigault et al.(2018)]{rigault2018} Rigault, M., Brinnel, V., Aldering, G., et al.\ 2018, arXiv e-prints, arXiv:1806.03849

\bibitem[Rigault et al.(2019)]{rigault2019} Rigault, M., Neill, J.~D., Blagorodnova, N., et al.\ 2019, \aap, 627, A115

\bibitem[Saunders et al.(2018)]{saunders2018} Saunders, C.,
  Aldering, G., Antilogus, P., et al.\ 2018, \apj, 869, 167

\bibitem[Sullivan et al.(2010)]{sullivan2010} Sullivan, M., Conley, A., Howell, D.~A., et al.\ 2010, \mnras, 406, 782

\bibitem[Wong et al.(2019)]{wong2019} Wong, K.~C., Suyu, S.~H., Chen, G.~C.-F., et al.\ 2019, arXiv e-prints, arXiv:1907.04869

\end{thebibliography}
%
\bibliographystyle{aa} 

\end{document}